\documentclass[footinbib,aps,prb,twocolumn,superscriptaddress,groupedaddress,a4paper]{revtex4}  
\usepackage{graphicx}
\usepackage{dcolumn}
\usepackage{bm}
\usepackage{amssymb}
\usepackage{amsmath}
\usepackage{subfigure}
\usepackage{color}
\begin{document}

\widetext
\title{Optical properties of metamorphic type-I InAs$_{1-x}$Sb$_{x}$/Al$_{y}$In$_{1-y}$As\\quantum wells grown on GaAs for the mid-infrared spectral range}


\author{Eva Repiso}
\email{e.repisomenendez1@lancaster.ac.uk} 
\affiliation{Physics Department, Lancaster University, Lancaster LA1 4YB, U.K.}

\author{Christopher A.~Broderick}
\email{c.broderick@umail.ucc.ie} 
\affiliation{Tyndall National Institute, Lee Maltings, Dyke Parade, Cork T12 R5CP, Ireland}
\affiliation{Department of Physics, University College Cork, Cork T12 YN60, Ireland}

\author{Maria de la Mata}
\affiliation{Facultad de Ciencias, IMEYMAT, Universidad de C\'{a}diz, C\'{a}diz, Spain}

\author{Reza Arkani}
\affiliation{Tyndall National Institute, Lee Maltings, Dyke Parade, Cork T12 R5CP, Ireland}
\affiliation{Department of Physics, University College Cork, Cork T12 YN60, Ireland}

\author{Qi Lu}
\affiliation{Physics Department, Lancaster University, Lancaster LA1 4YB, U.K.}

\author{Andrew R.~J.~Marshall}
\affiliation{Physics Department, Lancaster University, Lancaster LA1 4YB, U.K.}

\author{Sergio I.~Molina}
\affiliation{Facultad de Ciencias, IMEYMAT, Universidad de C\'{a}diz, C\'{a}diz, Spain}

\author{Eoin P.~O'Reilly}
\affiliation{Tyndall National Institute, Lee Maltings, Dyke Parade, Cork T12 R5CP, Ireland}
\affiliation{Department of Physics, University College Cork, Cork T12 YN60, Ireland}

\author{Peter J.~Carrington}
\affiliation{Department of Engineering, Lancaster University, Lancaster LA1 4YW, U.K.}

\author{Anthony Krier}
\affiliation{Physics Department, Lancaster University, Lancaster LA1 4YB, U.K.}

\date{\today}


\begin{abstract}

We analyse the optical properties of InAs$_{1-x}$Sb$_{x}$/Al$_{y}$In$_{1-y}$As quantum wells (QWs) grown by molecular beam epitaxy on relaxed Al$_{y}$In$_{1-y}$As metamorphic buffer layers (MBLs) using GaAs substrates. The use of Al$_{y}$In$_{1-y}$As MBLs allows for the growth of QWs having large type-I band offsets, and emission wavelengths $> 3$ $\mu$m. Photoluminescence (PL) measurements for QWs having Sb compositions up to $x = 10$\% demonstrate strong room temperature emission up to 3.4 $\mu$m, as well as enhancement of the PL intensity with increasing wavelength. To quantify the trends in the measured PL we calculate the QW spontaneous emission, using a theoretical model based on an 8-band \textbf{k}$\cdot$\textbf{p} Hamiltonian. The theoretical calculations, which are in good agreement with experiment, identify that the observed enhancement in PL intensity with increasing wavelength is associated with the impact of compressive strain on the QW valence band structure. Our results highlight the potential of type-I InAs$_{1-x}$Sb$_{x}$/Al$_{y}$In$_{1-y}$As metamorphic QWs to address several limitations associated with existing heterostructures operating in the mid-infrared, establishing these novel heterostructures as a suitable platform for the development of mid-infrared light-emitting diodes.

\end{abstract}


\maketitle


There is increasing interest in the development of compact and inexpensive semiconductor light sources operating at mid-infrared wavelengths between 2 and 5 $\mu$m, due to their potential for a wide variety of sensing applications, including monitoring of atomospheric pollutants, chemical process control, and detection of biological markers in non-invasive medical diagnostics, in addition to potential applications in free-space optical communications. \cite{Krier_book_2006,Bauer_SST_2011,Hodgkinson_MST_2013,Jung_JO_2017} Due to the presence of strong absorption features in the vibrational-rotational spectra of the important greenhouse gases methane (CH$_{4}$) and carbon dioxide (CO$_{2}$) at respective wavelengths of 3.3 and 4.2 $\mu$m, devices operating at these wavelengths are of particular interest for environmental monitoring. \cite{Hodgkinson_MST_2013}

Over the past decade, significant advances have been made in the development of GaSb-based diode lasers and light emitting diodes (LEDs). \cite{Vizbaras_EL_2011,Vizbaras_SST_2012,Sifferman_JSTQE_2015} In the 2 -- 3 $\mu $m spectral range type-I GaInAsSb/AlGa(In)AsSb QWs have demonstrated impressive characteristics, but their performance at and above room temperature degrades significantly for wavelengths $\lambda \gtrsim 3$ $\mu$m due to a combination of Auger recombination and thermal leakage of holes. \cite{Sheterengas_SST_2004,Brien_APL_2006,Tossou_SST_2013,Eales_JSTQE_2017} Further limitations to achieving $\lambda \gtrsim 3$ $\mu$m in GaSb-based heterostructures relate to the presence of (i) a miscibility gap in In- and As-rich GaInAsSb alloys, leading to a reduction in material quality, and (ii) a band structure in which the valence band (VB) spin-orbit splitting energy is close in magnitude to the band gap, leading to increased hot-hole producing (CHSH) Auger recombination and inter-valence band absorption.

While inter- and intra-band cascade devices have become well established at wavelengths above 3 $\mu$m, \cite{Bismuto_SST_2012,Tournie_SS_2012,Vurgaftman_JSTQE_2013,Vurgaftman_JPDAP_2015,Vitiello_OE_2015,Razeghi_OE_2015} these are complicated structures requiring careful design and optimisation: a typical cascade active region consists of up to 100 layers with tight tolerances on thickness and composition, with the requirement for reproducibility then placing strong demands on epitaxial growth. Furthermore, due to the relative expense and technological immaturity of the GaSb and InAs platforms compared to the GaAs or InP platforms commonly employed in near-infrared optical communications, it is preferable to develop mid-infrared devices on either GaAs or InP substrates. Doing so has the potential to reduce fabrication costs, as well as to allow to take advantage of the existing array of high-performance GaAs- and InP-based passive photonic components and integrated circuitry.


\begin{figure*}[t!]
	\includegraphics[width=1.00\textwidth]{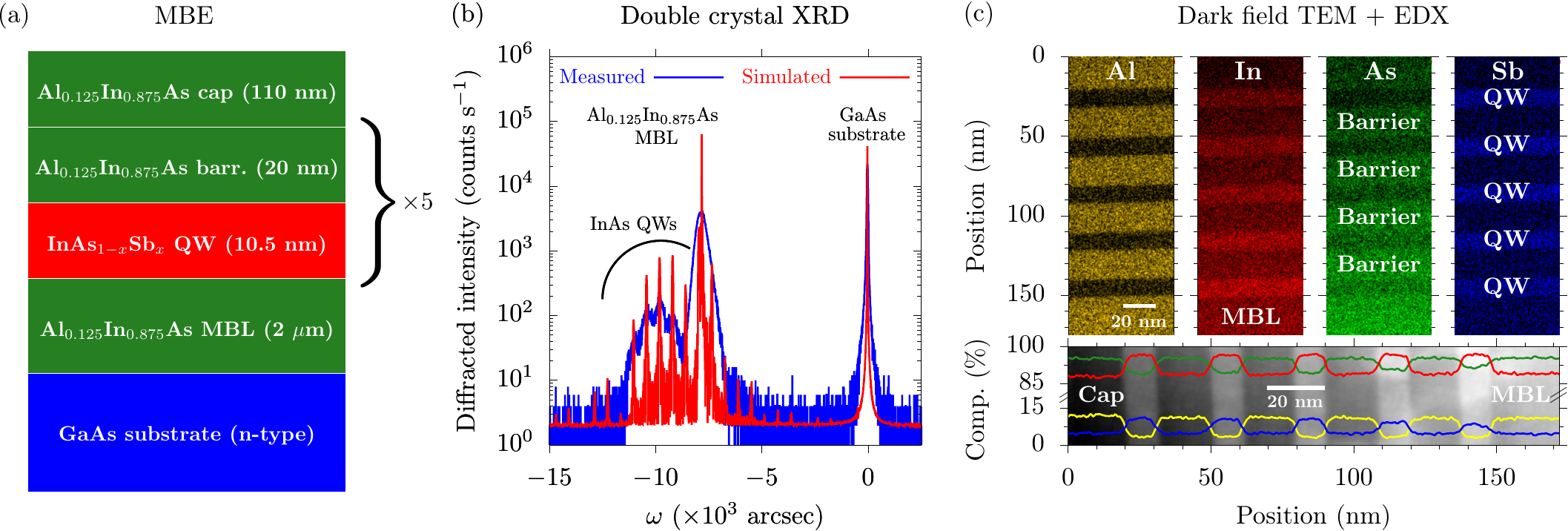}
	\caption{(a) Schematic illustration of the metamorphic multi-QW structures investigated in this work: the structures consist of compressively strained InAs$_{1-x}$Sb$_{x}$ QWs grown between unstrained Al$_{0.125}$In$_{0.875}$As barriers, with growth on GaAs facilitated via the use of an Al$_{0.125}$In$_{0.875}$As MBL. (b) DC-XRD $\omega - 2 \theta$ rocking curve for MQW1, measured (solid blue line) and simulated (solid red line) about the GaAs (004) reflection. (c) Upper panels: false-colour dark-field TEM images for MQW4, where the intensity of the yellow, red, green and blue colouration respectively describes the Al, In, As and Sb compositions in the QW and barrier layers. Lower panel: solid yellow, red, green and blue lines respectively denote the variation of the Al, In, As and Sb compositions along the (001) growth direction for MQW4, as inferred from EDX measurements. For illustrative purposes, the EDX data have been superimposed onto a dark-field TEM image of the structure.}
	\label{fig:structures}
\end{figure*}

Here, we present and analyse InAs$_{1-x}$Sb$_{x}$/Al$_{y}$In$_{1-y}$As quantum wells (QWs) grown on GaAs substrates, where growth of these lattice-mismatched heterostructures is facilitated via relaxed Al$_{y}$In$_{1-y}$As metamorphic buffer layers (MBLs). These QWs offer deep type-I band offsets, providing good confinement of both electrons and holes, and hence respectively maximising and minimising the electron-hole spatial overlap and thermal carrier leakage. The prototypical structures we investigate demonstrate (i) strong room temperature photoluminescence (PL) up to $\lambda = 3.4$ $\mu$m, and (ii) a steady increase in peak and integrated PL intensity with increasing emission wavelength. Using theoretical calculations we identify and quantify the origin of this behaviour, highlighting the key role played by compressive strain in the QW layers. Moreover, we describe general trends in the electronic and optical properties of these novel heterostructures, and on this basis evaluate their potential for applications in mid-infrared LEDs. Our analysis demonstrates that the characteristics of these metamorphic QWs are promising for the development of LEDs operating in the 3 -- 4 $\mu$m wavelength range, and in particular for sensing applications at wavelengths close to 3.3 $\mu$m.

The remainder of this paper is organised as follows. Firstly, we describe the growth and characterisation of the structures investigated, and the experimental measurements of the optical properties. Secondly, we describe the theoretical model used to analyse the electronic and optical properties. We then present our experimental and theoretical results, beginning with an analysis of general properties of metamorphic InAs$_{1-x}$Sb$_{x}$/Al$_{y}$In$_{1-y}$As QWs, before analysing the grown structures. Finally, we summarise and conclude.


The structures investigated were grown on (001)-oriented n-type GaAs substrates, using a Veeco GENxplor molecular beam epitaxy (MBE) system. Valved cracker cells were used to provide the As$_{2}$ and Sb$_{2}$ fluxes, while thermal effusion K-cells were used to provide the In and Al fluxes. In-situ reflection high-energy electron diffraction was used to monitor surface reconstruction. A schematic illustration of the MBE-grown structures is shown in Fig.~\ref{fig:structures}(a). Each structure consists of a 0.4 $\mu$m thick GaAs buffer, grown at 570 $^{\circ}$C, atop which a 2 $\mu$m thick relaxed Al$_{0.125}$In$_{0.875}$As MBL was grown at 510 $^{\circ}$C. The structures were cooled to 450 $^{\circ}$C for the growth of the five-period InAs$_{1-x}$Sb$_{x}$/Al$_{0.125}$In$_{0.875}$As active multi-QW layers. The respective thicknesses of the InAs$_{1-x}$Sb$_{x}$ QW and Al$_{0.125}$In$_{0.875}$As barrier layers were 10.5 and 20 nm. Finally, a 110 nm thick Al$_{0.125}$In$_{0.875}$As cap layer was deposited at a temperature of 450 $^{\circ}$C.

The structures were characterised by a combination of double-crystal x-ray diffraction (DC-XRD), atomic force microscopy and transmission electron microscopy (TEM). The solid blue line in Fig.~\ref{fig:structures}(b) shows the DC-XRD $\omega - 2 \theta$ rocking curve for structure MQW1, measured about the GaAs (004) reflection. The solid red line shows the corresponding simulated rocking curve, obtained using the Bede RADs software. We observe good overall correspondence between the measured and simulated rocking curves, confirming in particular that the Al$_{0.125}$In$_{0.875}$As MBL and InAs$_{1-x}$Sb$_{x}$ QWs are respectively fully relaxed and pseudomorphically strained. The full-width at half-maximum of the measured peak in the DC-XRD rocking curve corresponding to the relaxed Al$_{0.125}$In$_{0.875}$As MBL is 759 arcsec, which compares favourably with previous reports. \cite{Jiang_ASS_2008,Choi_JKPS_2009,Cao_APL_2013} Several satellite peaks associated with the compressively strained InAs$_{1-x}$Sb$_{x}$ QWs are observed in the measured and simulated rocking curves, from which the QW strain, and hence Sb composition $x$, can be estimated (cf.~Table~\ref{tab:structures}). For the four structures MQW1, MQW2, MQW3 and MQW4 grown, we estimate respective QW Sb compositions $x = 0$, 3, 6 and 10\%.


\begin{table*}[t!]
	\caption{\label{tab:structures} Details of the MBE-grown InAs$_{1-x}$Sb$_{x}$/Al$_{y}$In$_{1-y}$As multi-QW structures studied, including the nominal QW Sb composition $x$, measured PL peak energies at $T = 4$ and 300 K, calculated in-plane strain $\epsilon_{xx}$, and calculated CB, HH and LH type-I band offsets $\Delta E_{\protect\scalebox{0.7}{\textrm{CB}}}$, $\Delta E_{\protect\scalebox{0.7}{\textrm{HH}}}$ and $\Delta E_{\protect\scalebox{0.7}{\textrm{LH}}}$ at $T = 300$ K. Each structure contains five InAs$_{1-x}$Sb$_{x}$ QWs, of nominal thickness 10.5 nm. Sb compositions in parentheses are those determined using the theoretical model to fit to the measured PL peak energy at $T = 300$ K. The PL peak energies in parentheses are the corresponding calculated SE peak energies at $T = 4$ and 300 K using the best-fit Sb compositions.}
	\begin{ruledtabular}
		\begin{tabular}{cccccccc}
			Structure & $x$ (\%) & PL peak, 4 K (eV) & PL peak, 300 K (eV) & $\epsilon_{xx}$ (\%) & $\Delta E_{\scalebox{0.7}{\textrm{CB}}}$ (meV) & $\Delta E_{\scalebox{0.7}{\textrm{HH}}}$ (meV) & $\Delta E_{\scalebox{0.7}{\textrm{LH}}}$ (meV) \\
			\hline
			MQW1 & 0.0  (0.0) & $0.495$ ($0.480$) & $0.437$ ($0.426$) & $-0.82$ & $151$ & $89$  & $33$  \\
			MQW2 & 3.0  (4.2) & $0.448$ ($0.438$) & $0.395$ ($0.384$) & $-1.03$ & $133$ & $149$ & $79$  \\
			MQW3 & 6.0  (5.5) & $0.457$ ($0.427$) & $0.383$ ($0.372$) & $-1.23$ & $120$ & $170$ & $89$  \\
			MQW4 & 10.0 (7.0) & $0.420$ ($0.414$) & $0.370$ ($0.359$) & $-1.51$ & $104$ & $197$ & $100$ \\
		\end{tabular}
	\end{ruledtabular}
\end{table*}

TEM experiments were performed to evaluate the quality of MQW4, since the QWs in this structure -- which have the highest QW Sb composition $x$ -- are most prone to Sb segregation. We focus here on the TEM analysis of the active (multi-QW) region. For reference, low magnification scanning TEM images of the entire structure of MQW4 are provided as Supplementary Material, where we note high crystalline and structural quality of the Al$_{y}$In$_{1-y}$As MBL, as well as the overall multi-QW structure. The upper panel of Fig.~\ref{fig:structures}(c) shows the spatial distribution of the chemical constituents Al, In, As and Sb, within the active region of structure MQW4. The compositions mapped via energy dispersive x-ray spectroscopy (EDX) performed in scanning TEM mode are displayed in yellow, red, green and blue respectively. The bottom panel of Fig.~\ref{fig:structures}(c) shows the alloy composition along the growth direction extracted from the EDX measurements, with the line colours for Al, In, As and Sb corresponding to those in the upper panels. The TEM results evidence the formation of pristine InAs$_{1-x}$Sb$_{x}$ QWs having sharp interfaces with the Al$_{0.125}$In$_{0.875}$As barriers, with only minor variations in QW thickness. The EDX data for the active region indicate average Sb compositions $x$ between 10.0 and 11.5\% (solid blue line) in the QW layers. We note that the Sb composition tends to increase in each QW along the (001) growth direction -- i.e.~the bottom QW has an average Sb composition $x \approx 10$\%, while the average Sb composition of the QW closest to the surface is approximately 11.5\%. For group-III elements, the barrier Al composition $y$ (solid yellow line) tends also to be slightly higher in the Al$_{y}$In$_{1-y}$As layers of the structure closest to the surface, with a measured average value close to $y \approx 11$\% in the barrier on the underside of the first QW being within 2\% of the average value $y \approx 12.5$\% in the barrier on the topside of the fifth QW (i.e.~in the capping layer). The TEM and EDX analyses then demonstrate that high crystalline quality is achieved for both the InAs$_{1-x}$Sb$_{x}$ QW and Al$_{y}$In$_{1-y}$As barrier layers of the structure (cf.~Fig.~\ref{fig:structures}(a)), displaying coherent and abrupt interfaces. Overall, these chemical analyses reveal a slight modulation of the QW Sb composition, which tends to increase towards the surface of the structure (cf.~Fig.~\ref{fig:structures}(c)), representing the most significant non-uniformity present in these MBE-grown structures.


To analyse the optical properties of the structures described in Table~\ref{tab:structures} we have performed temperature-dependent photoluminescence (PL) measurements. The PL measurements were carried out using a 785 nm diode-pumped solid state laser as the optical (excitation) source, with a continuous-wave output power of 200 mW, while the sample temperature was varied between $T = 4$ and 300 K by means of a closed-cycle He cryostat. The radiation emitted from the structures upon excitation was collected using CaF$_{2}$ lenses and focused onto a Fourier transform mid-infrared spectrometer.


Our theoretical calculations of the electronic and optical properties of these structures were based on an 8-band \textbf{k}$\cdot$\textbf{p} Hamiltonian, \cite{Bahder_PRB_1990} implemented for QW heterostructures via a numerically efficient reciprocal space (plane wave) method. \cite{Healy_JQE_2006,Ehrhardt_book_2014} The temperature dependence of the InAs$_{1-x}$Sb$_{x}$ and Al$_{y}$In$_{1-y}$As band gaps is described via a conventional Varshni parametrisation. \cite{Varshni_P_1967,Vurgaftman_JAP_2001} The band structure and eigenstates obtained via the multi-band \textbf{k}$\cdot$\textbf{p} calculation for a given QW structure are used directly to compute spontaneous emission (SE) spectra, under the assumption of quasi-equilibrium (thermal) carrier distributions. To facilitate comparison to experiment we perform SE calculations at a fixed carrier density $n = 10^{15}$ cm$^{-3}$ per QW, a low value selected to replicate a typical carrier density associated with optical excitation in the PL measurements. Full details of the theoretical model -- which is based upon that we have previously developed to analyse the properties of 1.3 $\mu$m metamorphic QW lasers, as well as near- and mid-infrared dilute bismide QW lasers \cite{Bogusevschi_IEEEJQE_2016,Broderick_IEEEJSTQE_2015,Marko_SR_2016,Broderick_SST_2018} -- will be presented in Ref.~\onlinecite{Arkani_submitted_2018}. We assume ideal, compressively strained InAs$_{1-x}$Sb$_{x}$ QWs of thickness 10.5 nm, surrounded by unstrained Al$_{0.125}$In$_{0.875}$As barriers. For the Sb-free ($x = 0$) structure MQW1 we note good agreement between theory and experiment, with only an 11 meV difference between the measured PL and calculated SE peak energies at room temperature. For MQW2 -- MQW4 the QW lattice constants inferred from the XRD measurements are used directly to compute the strain in the QW layers. Since our focus here is on the evolution of the optical properties with room temperature emission wavelength, for the theoretical calculations the QW Sb composition $x$ is varied to reproduce -- in the calculated SE spectrum for each structure -- the shift in the measured room temperature PL peak energy relative to MQW1. The Sb compositions obtained in this manner are given in parentheses in Table~\ref{tab:structures}. The best-fit value of $x$ is close to the nominal Sb composition indicated by experimental characterisation for MQW3, but is higher (lower) than the nominal value in MQW2 (MQW4). A number of factors may contribute to discrepancies in these theoretical best-fit Sb compositions, including fluctations in QW alloy composition and thickness (cf.~Fig.~\ref{fig:structures}(c)). For simplicity we refer throughout the text to the nominal, rather than theoretical best-fit, Sb compositions.


\begin{figure*}[t!]
	\includegraphics[width=1.00\textwidth]{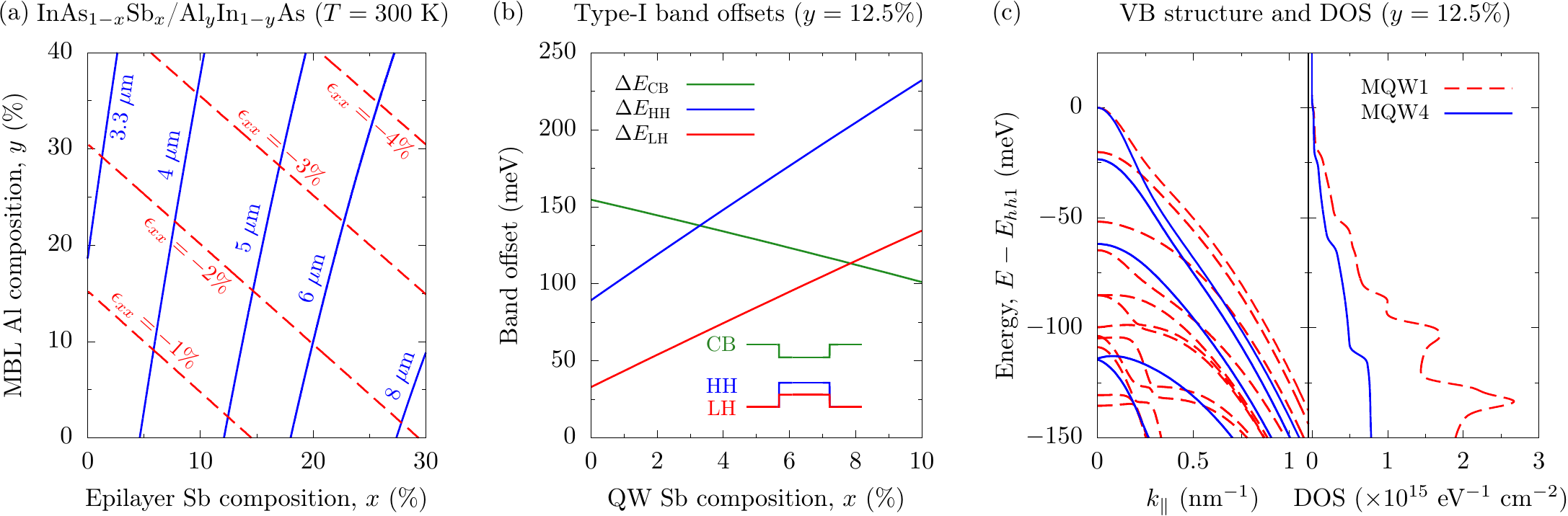}
	\caption{(a) Composition space map describing the ranges of in-plane strain ($\epsilon_{xx}$) and band gap (at $T = 300$ K) accessible using bulk-like InAs$_{1-x}$Sb$_{x}$ epitaxial layers grown on an Al$_{y}$In$_{1-y}$As MBL. Dashed red and solid blue lines respectively denote paths in the composition space along which $\epsilon_{xx}$ and the band gap are constant. (b) Calculated variation of the CB, HH and LH type-I band offsets (solid green, blue and red lines, respectively) with Sb composition $x$, for compressively strained InAs$_{1-x}$Sb$_{x}$ QWs grown on an Al$_{0.125}$In$_{0.875}$As MBL. (c) Calculated VB structure (left panel) and DOS (right panel) for structures MQW1 (dashed red lines) and MQW4 (solid blue lines) of Table~\ref{tab:structures}. For each structure the zero of energy has been set at the QW VB edge -- i.e.~at the energy of the highest energy valence subband $hh1$, which is purely HH-like at $k_{\parallel} = 0$.}
	\label{fig:theory}
\end{figure*}




We begin our analysis by considering the calculated band structure of bulk-like InAs$_{1-x}$Sb$_{x}$ epitaxial layers grown on Al$_{y}$In$_{1-y}$As MBLs. The solid blue (dashed red) lines in Fig.~\ref{fig:theory}(a) denote compositions for which the band gap $E_{g}$ (in-plane strain $\epsilon_{xx}$) is constant. The lattice constant of the Al$_{y}$In$_{1-y}$As MBL varies between that of InAs and AlAs, equalling that of InP for $y = 47.7$\%. Figure~\ref{fig:theory}(a) suggests that (i) $E_{g} = 0.376$ eV ($\lambda = 3.3$ $\mu$m) can be achieved, e.g., in bulk for $x \lesssim 5$\% but with large MBL Al compositions $y \gtrsim 20$\%, corresponding to large compressive strains $\epsilon_{xx} \approx -2$\%, and (ii) $E_{g} = 0.248$ eV ($\lambda = 5$ $\mu$m) can be achieved, e.g., at $x \approx 14$\% ($\epsilon_{xx} \approx -1.5$\%) via growth on an Al$_{0.1}$In$_{0.9}$As MBL.

These results suggest that compressively strained InAs$_{1-x}$Sb$_{x}$ epitaxial layers grown on relaxed Al$_{y}$In$_{1-y}$As MBLs offer access to a broad range of mid-infrared emission wavelengths $\lambda \gtrsim 3$ $\mu$m. We note however that the calculations presented in Fig.~\ref{fig:theory}(a) are for bulk-like epitaxial layers. When considering QWs the confinement energy -- totalling approximately 50 meV for the lowest energy bound CB ($e1$) and highest energy bound hole ($h1$) states in full numerical calculations -- must be accounted for. Achieving a desired QW $e1$-$h1$ band gap then requires growth of InAs$_{1-x}$Sb$_{x}$ QWs having a bulk band gap which is lower by approximately 50 meV, requiring a significant increase in $x$ and hence $\vert \epsilon_{xx} \vert$ and thereby limiting the accessible wavelength range. Using the expression due to Voisin, \cite{Voisin_SPIE_1988,Reilly_SST_1989} we estimate a strain-thickness limit $t_{c} \vert \epsilon_{xx} \vert \approx 23$ nm \% for InAs$_{1-x}$Sb$_{x}$/Al$_{0.125}$In$_{0.875}$As. This suggests a maximum permissible strain $\vert \epsilon_{xx} \vert \approx 2.3$\% for a 10 nm thick QW in the structures grown and analysed here, corresponding to $x \approx 22$\%. For such a InAs$_{0.78}$Sb$_{0.22}$/Al$_{0.125}$In$_{0.875}$As QW we calculate an $e1$-$h1$ transition energy of 0.221 eV at $T = 300$ K, suggesting that the room temperature emission wavelength is restricted to $\lesssim 5.6$ $\mu$m by strain-thickness limitations. Further analysis (below) suggests that this estimated upper limit on the emission wavelength is likely reduced in practice.


\begin{figure*}[t!]
	\includegraphics[width=1.00\textwidth]{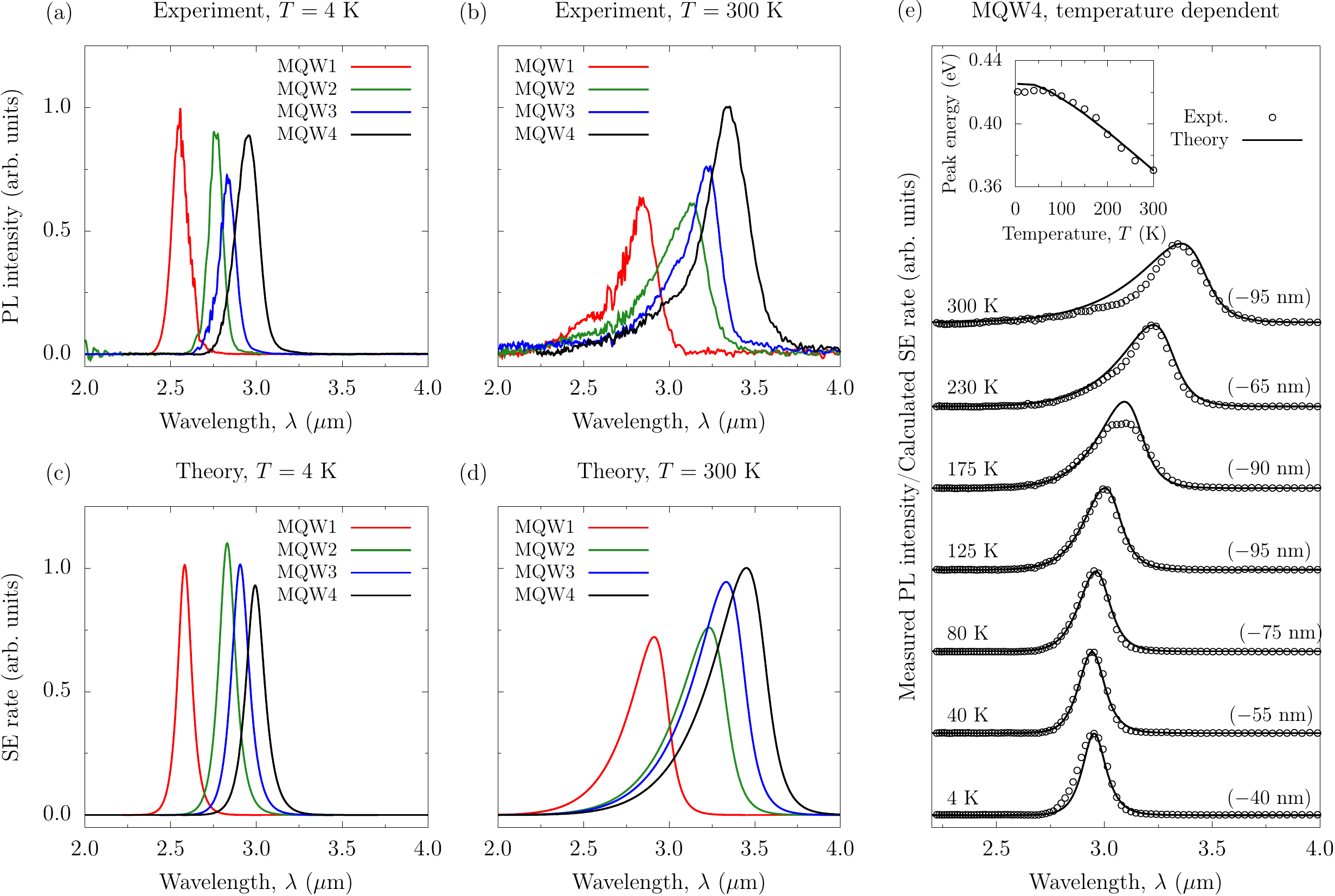}
	\caption{Measured PL spectra for structures MQW1 -- MQW4 (solid red, green, blue and black lines, respectively) at temperatures (a) $T = 4$ K, and (b) $T = 300$ K. The corresponding calculated SE spectra for structures MQW1 -- MQW4 at $T = 4$ and 300 K are shown respectively in (c) and (d). At each temperature the values of the measured PL and calculated SE have been normalised to that of the structure having the highest measured PL intensity (MQW1 at 4 K, and MQW4 at 300 K). The full temperature-dependent set of measured PL and calculated SE spectra for MQW4 are shown in (e) using, respectively, open circles and solid black lines. The calculated SE spectra in (e) have, at each temperature, been normalised in intensity (except at $T = 175$ K; see text) and shifted in wavelength (by the amount specified in parentheses) to match the experimental data at the PL peak. The open circles (solid black line) in the inset to (e) show the measured (calculated) variation of the PL (SE) peak energy with temperature in MQW4, where a rigid 11 meV redshift has been applied to the calculated SE peak energies.}
	\label{fig:theory_vs_expt}
\end{figure*}


Turning our attention now to the InAs$_{1-x}$Sb$_{x}$/Al$_{0.125}$In$_{0.875}$As QWs of interest here, the solid green, blue and red lines in Fig.~\ref{fig:theory}(b) respectively show the calculated type-I conduction band (CB), heavy-hole (HH) and light-hole (LH) QW band offsets $\Delta E_{\scalebox{0.6}{\textrm{CB}}}$, $\Delta E_{\scalebox{0.6}{\textrm{HH}}}$ and $\Delta E_{\scalebox{0.6}{\textrm{LH}}}$, as a function of the Sb composition $x$ in the QW. At $x = 0$\% (MQW1) we calculate large type-I CB and HH band offsets, $\Delta E_{\scalebox{0.6}{\textrm{CB}}} = 151$ meV and $\Delta E_{\scalebox{0.6}{\textrm{HH}}} = 89$ meV (cf.~Table~\ref{tab:structures}). As $x$ increases we calculate a strong increase in $\Delta E_{\scalebox{0.6}{\textrm{HH}}}$, to 232 meV at $x = 10$\%, as a result of the associated increases in (i) the VB offset, and (ii) the magnitude $\vert \epsilon_{xx} \vert$ of the compressive strain. Conversely, $\Delta E_{\scalebox{0.6}{\textrm{CB}}}$ is calculated to decrease strongly with increasing $x$ -- to 97 meV at $x = 10$\% -- due to the upward shift of the InAs$_{1-x}$Sb$_{x}$ CB edge energy with increasing $\vert \epsilon_{xx} \vert$. As such, the ionisation energy for a bound $e1$ electron decreases strongly with increasing $x$, from 110 meV at $x = 0$\% to 64 meV at $x = 10$\%. This ionisation energy is calculated as the difference between the barrier CB edge and the energy of the bound $e1$ eigenstate at $k_{\parallel} = 0$, which is the maximum possible energy required to extract a bound electron from the QW, and at $x = 10$\% is only slightly larger than twice the average thermal energy $k_{\scalebox{0.6}{\textrm{B}}} T$ at room temperature. This analysis suggests that electron confinement is significantly degraded for $x \gtrsim 10$\%, corresponding in a full numerical calculation to $\lambda \gtrsim 3.9$ $\mu$m. We therefore conclude that favourable type-I band offsets -- providing high electron-hole spatial overlap and minimising thermal carrier leakage -- can be achieved for $\lambda \lesssim 4$ $\mu$m in InAs$_{1-x}$Sb$_{x}$/Al$_{0.125}$In$_{0.875}$As QWs having $x \lesssim 10$\%. Extending the emission wavelength beyond 4 $\mu$m is then likely to require careful QW design and optimisation. \cite{Arkani_NUSOD_2018,Arkani_submitted_2018}


The left- and right-hand panels of Fig.~\ref{fig:theory}(c) respectively show the calculated VB structure and density of states (DOS) for structures MQW1 (dashed red lines) and MQW4 (solid blue lines) of Table~\ref{tab:structures}. The larger compressive strain in MQW4 leads to reduced in-plane hole effective masses, and hence to an overall reduction in the VB edge DOS. On the basis of the calculated electronic properties we note that (i) achieving longer emission wavelengths requires higher Sb compositions $x$, but comes at the expense of degrading the electron confinement (cf.~Fig.~\ref{fig:theory}(b)), and (ii) for fixed MBL Al composition $y$ larger Sb compositions $x$ are associated with larger $\vert \epsilon_{xx} \vert$, which can be expected to reduce the VB edge DOS and enhance the radiative efficiency. \cite{Reilly_IEEEJQE_1994,Adams_IEEEJSTQE_2011} Our analysis therefore identifies an important trade-off for the design of optimised structures to target specific emission wavelengths: sufficiently high Sb compositions $x$ should be sought to generate appreciable compressive strain in the QW(s), while ensuring that $x$ is not sufficiently large that thermal leakage of electrons degrades the overall efficiency of a QW-LED device at and above room temperature. \cite{Arkani_submitted_2018}


Having described general trends in the calculated electronic properties of these structures, we turn our attention now to the optical properties. Figures~\ref{fig:theory_vs_expt}(a) and~\ref{fig:theory_vs_expt}(b) show the measured PL spectra for the structures of Table~\ref{tab:structures}, at respective temperatures $T = 4$ and 300 K. In each case the measured PL spectrum for MQW1 -- MQW4 is shown using solid red, green, blue and black lines, respectively. At each temperature the PL spectra have been normalised to the intensity of the structure having the highest measured peak PL intensity (MQW1 at $T = 4$ K, and MQW4 at $T = 300$ K). Examining first the PL spectra measured at $T = 4$ K in Fig.~\ref{fig:theory_vs_expt}(a), we note that incorporating Sb leads to a significant redshift of the emission wavelength. For the Sb-free structure MQW1 we measure a PL peak wavelength of 2.51 $\mu$m (0.495 eV), which shifts to 2.95 $\mu$m (0.420 eV) in MQW4 which contains 10\% Sb. At $T = 300$ K the temperature-induced band gap reduction redshifts these PL peak wavelengths to 2.84 and 3.35 $\mu$m (0.437 and 0.370 eV) respectively (cf.~Table~\ref{tab:structures}). For MQW1 we calculate respective peak PL wavelengths of 2.58 and 2.91 $\mu$m at $T = 4$ and 300 K, corresponding to respective PL peak energies of 0.480 and 0.426 eV. The threoretical calculations then underestimate the measured PL peak energy by only 15 meV (11 meV) at $T = 4$ K (300 K), suggesting good quantitative agreement between theory and experiment. We note that these small deviations between the measured and calculated values for the Sb-free structure are likely attibutable to inhomogeneities in the structure due, e.g., to fluctations in QW thickness, intermixing of atoms between the barrier and adjacent QW layers during growth, or uncertainties in the material parameters employed in the theoretical calculations.

At $T = 4$ and 300 K we observe that the measured PL spectra for the structures of Table~\ref{tab:structures} have similar overall character. Firstly, all spectra possess a single emission peak, which the theoretical calculations confirm as resulting from recombination involving electrons occupying states in the lowest energy ($e1$) QW conduction subband. This is consistent with the low carrier densities generated by optical excitation, upon which basis we expect that excited electrons thermalise rapidly into the $e1$ subband. While theoretical calculations at $T = 4$ K suggest that injected holes solely occupy states in the highest energy ($hh1$) valence subband, at $T = 300$ K a substantial fraction ($\approx 40$\%) occupy states in the second valence subband ($hh2$). However, the matrix elements for optical transitions between $e1$ and $hh2$ subband states vanish at the centre of the QW Brillouin zone ($k_{\parallel} = 0$), and remain small compared to those associated with $e1$--$hh1$ transitions at non-zero in-plane wave vector $k_{\parallel}$, indicating that recombination of $e1$ electrons with $hh2$ holes does not contribute appreciably to the optical emission. Secondly, while the measured PL spectra at $T = 4$ K are largely symmetric about the emission peak, at $T = 300$ K we note the presence of a pronounced high energy tail. This is consistent with the presence of thermalised carrier distributions, describable by separate electron and hole quasi-Fermi distribution functions, whereby carriers at higher temperature occupy subband states over larger ranges of energies than at low temperature.

Turning our attention to the intensity of the measured PL spectra, at $T = 4$ K we note that the measured peak PL intensity remains approximately constant for all structures, reducing by approximately 10\% in going from the Sb-free MQW1, to MQW4 in which $x = 10$\%. Conversely, at $T = 300$ K we note that the measured peak PL intensity increases strongly with increasing Sb composition $x$ -- or, equivalently, increasing emission wavelength -- with the measured peak PL intensity for MQW4 at $T = 300$ K being larger by approximately two-thirds than that measured for MQW1 at the same temperature. We note that the measured enhancement at room temperature of the peak PL intensity under illumination at fixed excitation power is promising from the perspective of device applications at wavelengths $\gtrsim 3$ $\mu$m, since theoretical analysis (below) indicates that the requirement to incorporate Sb to reach these wavelengths is associated with an intrinsic enhancement of the radiative efficiency of a multi-QW structure.


Figures~\ref{fig:theory_vs_expt}(c) and~\ref{fig:theory_vs_expt}(d) show the calculated SE spectra for the structures of Table~\ref{tab:structures}, at respective temperatures $T = 4$ and 300 K. The line colours and normalisation of the peak SE rate are as for the measured PL spectra in Figs.~\ref{fig:theory_vs_expt}(a) and~\ref{fig:theory_vs_expt}(b). We note good overall qualitative agreement between these calculated SE spectra and the measured PL spectra of Figs.~\ref{fig:theory_vs_expt}(a) and~\ref{fig:theory_vs_expt}(b). At $T = 4$ K the measured peak PL intensity and calculated peak SE rate reduce by $\lesssim 10$\% as $x$ increases from 0 (MQW1) to 10\% (MQW4), while at $T = 300$ K the peak measured PL intensity and calculated SE rate increases by $\gtrsim 40$\% between $x = 0$ and 10\%. On this basis we conclude that the theoretical model correctly captures the key observed trends in the experimental measurements: at low temperature the emission intensity at fixed injection is approximately independent of $x$ (or, equivalently, $\lambda$), while at room temperature the emission intensity increases strongly with increasing $\lambda$.

Figure~\ref{fig:theory_vs_expt}(e) shows the measured PL (open circles) and calculated SE (solid lines) spectra for MQW4 ($x = 10$\%), for a range of temperatures from $T = 4$ to 300 K. To facilitate comparison of the overall character of the measured and calculated spectra, the calculated SE spectra at each temperature have been (i) shifted in wavelength (by the amount denoted in parentheses in Fig.~\ref{fig:theory_vs_expt}(e)) to match the peak of the corresponding measured PL spectrum, and (ii) normalised to the measured peak PL intensity. This normalisation has not been carried out at $T = 175$ K due to the presence of what appears to be an absorption feature at $\lambda \approx 3.1$ $\mu$m, evident in the experimental data for $T \geq 175$ K and lying close to the PL peak at $T = 175$ K. We attribute this anomalous absorption, which is present at this wavelength in all of the samples studied, to the presence of atmospheric water (H$_{2}$O) vapour. Despite this, we again note good overall agreement between theory and experiment. The calculated SE spectra -- in which the spectral broadening is described theoretically by a hyperbolic secant lineshape \cite{Marko_SR_2016} of width $6.6$ meV -- describe that the spectral broadening of the measured PL spectra is typical of that of a conventional III-V semiconductor alloy, \cite{Tomic_IEEEJSTQE_2003} suggesting high material quality in the metamorphic QWs, in agreement with the results of the TEM measurements described above. The inset to Fig.~\ref{fig:theory_vs_expt}(e) shows the variation with temperature of the measured PL (open circles) and calculated SE (solid line) peak energy, where a rigid 11 meV redshift has been applied to the calculated data to account for the discrepancy between the measured PL and calculated SE peak energies at room temperature (cf.~Table~\ref{tab:structures}). The close agreement between theory and experiment here verifies that the temperature dependence of the measured PL peak energy is well described via the conventional Varshni parametrisation. \cite{Varshni_P_1967,Vurgaftman_JAP_2001} We note the presence of a weak ``s-shape'' temperature dependence of the measured PL peak energy for $T \lesssim 50$ K, which may be indicative of carrier localisation. \cite{Steenbergen_JL_2016} Weak carrier localisation of this nature would be compatible with, e.g., the observed minor, short-range fluctuations in the QW Sb composition and thickness observed via TEM imaging (cf.~Fig.~\ref{fig:structures}(c)), and is not sufficiently pronounced to be expected to impact device performance.

To identify and quantify the origin of the measured increase in room temperature PL intensity with increasing $\lambda$ we have analysed the distinct contributions -- the band gap (emission wavelength), inter-band optical matrix elements, and electron and hole quasi-Fermi levels \cite{Ahn_IEEEJQE_1990,Chang_IEEEJSTQE_1995,Tomic_IEEEJSTQE_2003} -- to the calculated SE spectra for each structure. For a QW of fixed thickness the SE rate at fixed carrier density $n$ is directly proportional to the photon energy, so that the decrease in band gap with increasing $x$ acts to decrease the peak SE rate. For MQW1 we calculate an inter-band optical transition strength 19.42 eV -- where the relevant scale \cite{Szmulowicz_PRB_1995} is the Kane parameter $E_{P}$ -- for the TE-polarised $e1$-$hh1$ transition at $k_{\parallel} = 0$, which decreases to 18.37 eV in MQW4. The changes of both the band gap and inter-band optical matrix elements with increasing $x$ then act to reduce the peak SE rate at fixed $n$, suggesting that the filling of the QW energy bands -- described by the quasi-Fermi levels, and determined primarily by the VB edge DOS -- is responsible for the calculated increase in the peak SE rate with increasing $\lambda$.

The calculated carrier (quasi-Fermi) distributions for MQW1 -- MQW4 confirm that the reduction of the VB edge DOS brought about by the increased compressive strain at larger $x$ (cf.~Fig.~\ref{fig:theory}(c)) is sufficient to increase the peak SE rate at fixed $n$. The larger effective masses of VB holes compared to those of CB electrons lead to holes occupying $hh1$ subband states over a larger range of in-plane wave vector $k_{\parallel}$ than those occupied by electrons in the $e1$ subband. The strict \textbf{k}-selection associated with optical transitions then renders holes occupying states at larger $k_{\parallel}$ unavailable to undergo radiative recombination, due to a lack of $e1$ electrons occupying states at equal $k_{\parallel}$. \cite{Adams_IEEEJSTQE_2011} As the in-plane effective mass of the $hh1$ VB decreases with increasing $x$, the DOS at the VB edge reduces to more closely match that at the CB edge. This allows holes to occupy $hh1$ subband states over a reduced range of $k_{\parallel}$ -- resulting in a reduction of the hole quasi-Fermi level at fixed $n$ -- making more electron-hole pairs available to undergo radiative recombination and contribute to the SE.

At $T = 4$ K this effect is not pronounced since, at low temperature, the quasi-Fermi distribution functions are step-like about the quasi-Fermi levels, with carriers occupying subband states across a limited range of $k_{\parallel}$. As a result, the calculated peak SE rate at $T = 4$ K depends only weakly on $x$ (cf.~Fig.~\ref{fig:theory_vs_expt}(c)). Conversely, at $T = 300$ K the hole quasi-Fermi distribution possesses a pronounced tail at energies below the hole quasi-Fermi level, describing that, on average, holes occupy states over a larger range of $k_{\parallel}$ than at low temperatures. As such, the impact of the strain-induced reduction in the VB edge DOS with increasing $x$ becomes pronounced, leading to the calculated 38\% increase in the peak SE rate between structures MQW1 and MQW4 at $T = 300$ K (cf.~Fig.~\ref{fig:theory_vs_expt}(d)).

To obtain a quantitative measure of the impact of Sb incorporation on the SE rate we have computed the radiative recombination coefficient $B$ at $T = 300$ K for each structure by (i) calculating the radiative current density $J_{\scalebox{0.7}{\textrm{rad}}}$ via integration of the calculated SE spectrum of Fig.~\ref{fig:theory_vs_expt}(d), and (ii) assuming $J_{\scalebox{0.7}{\textrm{rad}}} = e B n^{2}$ (i.e.~the Boltzmann approximation). Given the expected \cite{Hader_SPIE_2006} weak dependence of $B$ on $n$ up to $n \sim 10^{17}$ cm$^{-3}$ (a carrier density typical of operation of an electrically pumped QW-LED), we anticipate that the values of $B$ computed here at a low carrier density corresponding to optical excitation describe trends which should remain largely valid for practical LED devices. \cite{Arkani_submitted_2018} We calculate $B = 0.56$ and $0.77$ $\times 10^{-4}$ cm$^{2}$ s$^{-1}$ for MQW1 and MQW4 respectively, describing the predicted increase of approximately 38\% in the radiative emission rate between $x = 0$ and 10\%. More detailed theoretical analysis \cite{Arkani_NUSOD_2018,Arkani_submitted_2018} suggests that the radiative emission rate can be further enhanced in QWs emitting at 3.3 $\mu$m, via the design and optimisation of strain-balanced structures.

While the theoretical calculations above -- which consider only radiative recombination of carriers -- describe accurately the dependence of the measured peak PL energy on temperature, they do not quantitatively account for the observed decrease in the measured peak PL intensity with increasing temperature. For an ideal QW the radiative recombination coefficient $B$ varies inversely with temperature, so that an approximately 75-fold decrease in integrated emission intensity would be expected as the temperature is increased from 4 to 300 K. However, for the structures of Table~\ref{tab:structures} the measured peak PL intensity at $T = 300$ K is reduced by a factor of approximately 300 compared to that measured at $T = 4$ K. On this basis we conclude that non-radiative carrier recombination mechanisms -- most likely Auger recombination, given the relevant wavelength range -- play an important role in determining the properties and performance of these structures. While detailed investigation of non-radiative recombination in these structures is beyond the scope of the current work, we note on the basis of detailed analysis of the calculated electronic properties \cite{Arkani_submitted_2018} (i) that both the hot electron- and hole-producing CHCC and CHSH Auger recombination mechanisms are likely to play a role in limiting the overall radiative efficiency, and (ii) while it may be possible to engineer the QW VB structure to reduce the CHSH recombination rate, \cite{Broderick_SST_2012} the mixing of LH- and SO-like eigenstates brought about by strain and quantum confinement effects likely render the CHSH process thresholdless. \cite{Arkani_submitted_2018}


In summary, we have presented a combined experimental and theoretical analysis of the properties of metamorphic InAs$_{1-x}$Sb$_{x}$/Al$_{0.125}$In$_{0.875}$As QWs grown on GaAs substrates by MBE. Characterisation of a series of prototypical multi-QW structures via XRD and TEM demonstrated high material and structural quality. Using theoretical calculations we have quantified the potential offered by this new class of metamorphic heterostructures, highlighting that the ability to engineer compressively strained QWs possessing deep type-I band offsets makes these structures particularly attractive for applications in mid-infrared light-emitters. Experimental measurements for QWs having Sb compositions up to $x = 10$\% revealed (i) strong PL up to a wavelength of 3.4 $\mu$m at room temperature, beyond the maximum emission wavelengths demonstrated using related metamorphic structures grown on InP, and (ii) an increase in the measured peak PL intensity with increasing emission wavelength. Via detailed theoretical analysis of the SE from these structures we identified that this favourable behaviour is associated with the impact of compressive strain on the QW electronic properties, which can be expected in general to enhance radiative efficiency.

Overall, we conclude that InAs$_{1-x}$Sb$_{x}$/Al$_{y}$In$_{1-y}$As metamorphic QWs grown on GaAs substrates offer a promising route to realising LEDs displaying good performance in the application-rich but technologically challenging 3 -- 4 $\mu$m spectral range, with particular promise for sensing applications at wavelengths close to 3.3 $\mu$m. Further studies are now required both to quantify the loss mechanisms present in these structures, and to design optimised structures \cite{Arkani_NUSOD_2018,Arkani_submitted_2018} for practical applications.


\section*{Acknowledgements}

This work was supported by the European Commission via the Marie Sk\l{}odowska-Curie Innovative Training Network PROMIS [project no.~641899], Science Foundation Ireland [SFI; project no.~15/IA/3082], the Engineering and Physical Sciences Research Council, U.K.~[EPSRC; project no.~EP/N018605/1], the National University of Ireland [NUI; via the Post-Doctoral Fellowship in the Sciences, held by C.A.B.], the Royal Academy of Engineering [RAE; via Research Fellowship no.~10216/114, held by P.J.C.], the Spanish Ministry of Economy and Competitiveness [MINECO; project no.~TEC2017-86102-C2-2-R], the Andalusian Research Council [PAI; research group TEP-946 INNANOMAT]. UCA authors acknowledge co-financing from the European Union via the European Structural Funds. The authors thank Dr.~James A.~Keen (Lancaster University, U.K.) for useful discussions.






\end{document}